\documentclass[oneside]{article}
\usepackage{amsmath}
\usepackage{amssymb}
\usepackage{bbm}
\usepackage[margin=3cm]{geometry}
\usepackage{hyperref}
\usepackage{tikz}
\usepackage{braket}

\newtheorem{thm}{Theorem}[section]

\newtheorem{lemma}[thm]{Lemma}

\newtheorem{theorem}[thm]{Theorem}

\newtheorem{corollary}[thm]{Corollary}

\newcommand{\N}{\mathbb{N}}

\newcommand{\Z}{\mathbb{Z}}

\newcommand{\x}{\boldsymbol{x}}
\newcommand{\y}{\boldsymbol{y}}

\title{Rough Bounds for Emptiness Formation Probability in the 2d Dimer model using Reflection Positivity}
\author{Shannon Starr and Scott Williams\\
\small
University of Alabama at Birmingham, Applied Mathematics, Birmingham, AL 35294--1170\\
}

\date{October 20, 2018}

\begin{document}

\maketitle

\begin{abstract}
\setcounter{section}{0}
We summarize how to obtain rough bounds for one version of the emptiness formation probability in the 2d dimer model.
The methods we use are the same as have been developed to obtain EFP bounds in the 
1d XXZ model in a paper with Crawford, Ng and one of the authors. A main tool is reflection positivity for
the basic dimer model, as proved by Heilmann and Lieb.
We also state the corollary for a 1d quantum spin system which Suzuki showed is related to the 2d Ising model.
\end{abstract}

\section{Introduction} 

In a quantum spin system, the emptiness formation probability, henceforth abbreviated ``EFP,'' refers to the probability to find a large block of spins
all aligned, for example, all spins-up in a specified block.
These form simple models of some of the most basic correlation functions one could look for related to a quantum spin system.
Moreover, in a reference  \cite{KorepinEtAl},  Korepin and other physicists found that for some exactly-solvable models, there is interesting mathematics 
involved in the precise determination of the EFP quantity, including some information about the asymptotics
when the block-size becomes large.
As a topic of study Korepin initiated the interest in EFP.

Since then, the topic has taken on other angles, such as study of the EFP for two dimensional classical
statistical mechanics model, such as dimers \cite{ColomoPronkoSportiello}.
The main result of the present article is to show that a basic method allows for rough bounds
in the generalized context, as well.
The basic method is the method of reflection positivity \cite{FrohlichSimonSpencer,DysonLiebSimon,FrohlichLieb}
and it was previously used to bound EFP in the low-temperature-thermal- and ground-states of the XXZ quantum spin
system by Crawford, Ng and one of the present authors \cite{CrawfordNgStarr}.
Fortunately, there is a quantum spin system that has the analogous role in relation to dimers
that the XXZ model has for the six-vertex model.
This is an XY-type model discovered by Suzuki \cite{Suzuki}.
We will also state the consequence for that model, which is bounds for the staggered magnetization profile (i.e., the N\'eel ordering).

% The method of reflection positivity is known to be somewhat fragile. Just because it applies to one model, that
% does not guarantee that it also applies to a related model or a model with a small perturbation.
% (More precisely, models either do or do not possess the reflection positivity property. 
% And some perturbations
% do result in families of reflection positive models, but it is not generically the case.)
% However, it does seem to be more robust than, say, exact solvability.
% For example, most quantum spin systems that one thinks of as reflection positive have this property in all dimensions, not just d=1.
% Therefore, it is interesting to explore this technique even for models which are known
% to be exactly solvable, such as the 2d dimer problem.
% Let us state our results for that model, now.

\subsection{Results in the 2d dimer model}

As a set-up, given $N,M \in \N = \{1,2,\dots\}$, let $\mathbb{T}^2_{N,M}$ denote the 2d $N\times M$ torus.
This is the set of vertices
\begin{equation*}
    \mathbb{T}^2_{N,M}\, =\, \{(x,y) \in \Z^2\, :\, 0\leq x\leq N-1\, ,\ 0\leq y\leq M-1\}\, ,
\end{equation*}
The edge set is such that $(x,y)$ and $(x',y')$ are connected by an edge in the graph if 
and only if 
\begin{equation*}
    x \equiv x'+\delta_1\ (\operatorname{mod}\, N)\, ,\ \text{ and }\ 
    y \equiv y' + \delta_2\  (\operatorname{mod}\, M)\, ,\ \text{ for }\,
    (\delta_1,\delta_2) \in \{(1,0)\, ,\ (-1,0)\, ,\ (0,1)\, ,\ (0,-1)\}\, .
\end{equation*}
We will assume that $N$ is even.
The version of the set-up for the dimer model we will use is a four-point spin space
\begin{equation*}
    \Sigma\, =\, \{U,D,L,R\}
\end{equation*}
for those four symbols, and we will consider configurations to be
functions $\omega : \mathbb{T}^2_{N,M} \to \Sigma$.
But $\Omega_{N,M}$ will actually be the set of all such function $\omega$ satisfying the following 
matching conditions for all $(x,y) \in \mathbb{T}^2_{N,M}$:
\begin{itemize}
    \item if $\omega(x,y) = U$ then $\omega(x,[y+1]_M)=D$,
    \item if $\omega(x,y) = D$ then $\omega(x,[y-1]_M)=U$,
    \item if $\omega(x,y) = L$ then $\omega([x-1]_N,y)=R$,
    \item if $\omega(x,y) = R$ then $\omega([x+1]_N,y)=L$,
\end{itemize}
where, for all $a \in \Z$, we have $[a]_M$ is the point in $\{0,\dots,M-1\}$ congruent to $a$ modulo $M$,
and $[a]_N$ is the point in $\{0,\dots,N-1\}$ congruent to $a$ modulo $N$.

Because of these rules and periodic boundary conditions, the number of $U$'s is the same as the number of $D$'s
and the number of $L$'s is the same as the number of $R$'s.
The simplest thing one could consider is the uniform measure on $\Omega_{N,M}$.
(Note that $\Omega_{N,M}$ is non-empty because $N \in 2\N = \{2,4,\dots\}$ is even.)
Let us actually consider the next-simplest framework.
Given $z \in (0,\infty)$, let 
\begin{equation*}
    p_{N,M}(\omega;z)\, =\, z^{V_{N,M}(\omega)}\, ,\ \text{ where }\ V_{N,M}(\omega)\, =\, \sum_{(x,y) \in \mathbb{T}^2_{N,M}} \mathbf{1}_{\{U\}}(\omega(x,y))\, .
\end{equation*}
Then we let $Z_{N,M}(z) = \sum_{\omega \in \Omega_{N,M}} p_{N,M}(\omega;z)$ and we define
\begin{equation*}
    \mathbb{P}_{N,M}^{(z)}(\{\omega\})\, =\, \frac{p_{N,M}(\omega;z)}{Z_{N,M}(z)}\, .
\end{equation*}
This is the measure on $\Omega_{N,M}$ which assigns {\em fugacity} $z$ to vertical dimers (edges of $\mathbb{T}^2_{N,M}$, 
joining two
sites $(x,y)$ and $(x,[y+1]_M)$ with spins $U$ and $D$, respectively).

Given all of this we may state our first result as follows.

\begin{figure}
$$
\begin{tikzpicture}[xscale=0.3,yscale=0.3,white]
\foreach \y in {0,1,...,5}
{
\foreach \x in {11,...,22}
{
\draw[gray] (\x,\y) rectangle +(1,1);
}
}
\foreach \x in {11,13,...,19}
{
\draw[black,very thick] (\x,0) -- +(1,0) -- +(1,-1) -- +(2,-1) -- +(2,0);
}
\draw[black,very thick] (21,0) -- +(1,0) -- +(1,-1) -- +(2,-1);
\draw[gray] (23,-1) -- +(0,1);
\end{tikzpicture}
$$
\caption{The EFP event for the dimer model. On the first $2n$ squares of the bottom row, on the  even sublattice, the dimers are U, and on the odd sublattice they
are not.}
    \label{fig:FirstDimer}
\end{figure}
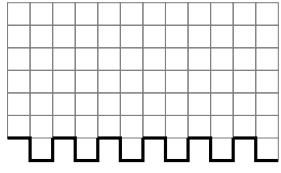

\begin{theorem}
\label{thm:main}
For each $n \in \N$, and $N \in 2\N \cap [2n,\infty)$, and any $M \in \N$, define an event
\begin{equation*}
    \mathcal{A}_{N,M}(n)\, =\, \left\{\omega \in \Omega_{N,M}\, :\, \forall x \in [0,2n)\cap \N\, ,\ \text{ we have } \big(\omega(x,0)=U\big) \Leftrightarrow  \big(x \in [0,2n) \cap (2\N)\big)\right\}\, .
\end{equation*}
Suppose $z \in (0,\infty)$.
Then there exist numbers $c,C \in (0,\infty)$ such that for each $n \in \N$ we have
\begin{equation*}
    \mathbb{P}_{N,M}^{(z)}\left(\mathcal{A}_{N,M}(n)\right)\in (e^{-Cn \cdot \min(n,M)},e^{-cn \cdot \min(n,M)})\, ,
\end{equation*}
as long as $N \in 2\N \cap [n,\infty)$. 
\end{theorem}

The event $\mathcal{A}_{N,M}(n)$ is stated in terms of the orientations of the dimers incident to just the first 1d row
of $\mathcal{T}^2_{N,M}$ (for $x$ between $0$ and $2n$).
But because of the dimer configuration constraints, it actually determines the configuration in a 2d diamond-shaped region:
\begin{lemma}
Suppose that $n \in \N$ and $N \in \N \cap [2n,\infty)$, and suppose that $M$ is in $\N \cap [2n,\infty)$.
Then for any $\omega \in \Omega_{N,M}$, we have that $\omega \in \mathcal{A}_{N,M}(n)$ if and only if
\begin{equation*}
\begin{split}
    &
    \forall y \in \N \cap (-n,n)\, ,\
    \forall x \in \N \cap [|y|,2n-2-|y|]\, ,\ \text{ we have }\\
    &\qquad \omega(x,y)=U \text{ if $x+y$ is even\, , }\ \omega(x,y)=D \text{ if $x+y$ is odd}\, .
\end{split}
\end{equation*}
\end{lemma}
This lemma is very easy to prove using induction.
Its significance is that it provides an intuitive motivation for why the probability of $\mathcal{A}_{N,M}(n)$
decays as the exponential of a constant times $n^2$ (assuming $M\geq n$) instead of simply decaying like the exponential of a constant times $n$.

\subsection{Implication for a Generalized XY Model of Suzuki}

In part in order to elucidate the relation between certain 1d quantum spin chains and 2d classical statistical mechanics
models, Suzuki introduced generalized XY models, which are also amenable to the Jordan-Wigner transform method, yielding
quadratic interactions in the resulting free Fermion creation and annihilation operators.
In particular, he found a generalized XY model related to the dimer problem \cite{Suzuki}.
Therefore, we may rephrase the result above using this.

Consider Suzuki's Hamiltonian (equivalent to the pure dimer model)
\begin{equation}
    H_{N}(z)\, =\, H^{(0)}_N + z^{-1} H^{(1)}_N\, ,
\end{equation}
where
\begin{equation}
    H^{(0)}_N\, =\, - \sum_{x=1}^N (\sigma_j^x \sigma_{j+1}^x - \sigma_j^y \sigma_{j+1}^y)\, ,
\end{equation}
and
\begin{equation}
    H^{(1)}_N\, =\,  \sum_{x=1}^N \sigma_j^z \left(1 + \frac{1}{2} \sigma_{j-1}^x \sigma_{j+1}^x + \frac{1}{2} \sigma_{j-1}^y \sigma_{j+1}^y\right)\, .
\end{equation}
Then we have similar bounds for the staggered magnetization profile of length $n$.

\begin{corollary}
For each $n \in \N$ and for all $N \in 2\N \cap [2n,\infty)$, we have that
\begin{equation*}
    \lim_{\beta \to \infty} \frac{\operatorname{Tr}\left[e^{-\beta H_N(z)} \prod_{j=1}^n \left(\frac{1}{2} + \frac{(-1)^j}{2} \sigma_j^z\right)\right]}
    {\operatorname{Tr}[e^{-\beta H_N(z)}]} \in (e^{-Cn^2},e^{-cn^2})\, ,
\end{equation*}
for the same constants $c$ and $C$ as in the theorem.
\end{corollary}
Note that in Suzuki's equivalence we have $\beta \to \infty$ for the quantum spin system, and we also have $M \to \infty$ in the dimer model.
We will not prove this corollary, here.
Suzuki proved the equivalence of the ground state for this Hamiltonian and 
the largest eigenvector of the row-to-row transfer matrix for the dimer problem explicated by Lieb  \cite{LiebDimer}.
Suzuki proceeded by generalizing the Jordan-Wigner transformation
and performing exact diagonalization.
Ideally, we would prefer an argument that uses the ``good signs condition,'' in the sense of the applicability of the Perron-Frobenius theorem.
But we are not sure this is the case for this Hamiltonian, and it will remain a question
that we hope is resolved in the future.

% The quantum spin system Hamiltonian discovered by Suzuki is actually equivalent to free Fermions. Suzuki pointed that out in his paper.
% But the methods do not actually use this fact, and in principle might be generalizable to other models which might not be as easily solved as non-interacting dimers.

\section{Reflection positivity and proof of Main Theorem}

In this note, we will indicate how to obtain upper bounds. In another, longer article, we plan to give a pedagogical guide to both the reflection positivity technique for obtaining
upper bounds as well as lower bounds. For now, we mention that lower bounds may be obtained by extending the technique of Crawford, Ng and one of the authors in \cite{CrawfordNgStarr}.

The dimer model is reflection positive, as proved by Heilmann and Lieb in \cite{HeilmannLieb}.
This implies several things:
firstly, if we reflect in the horizontal direction $k$ times, which results in $2^k$ copies of the original tile,
we have
\begin{equation}
    \mathbb{P}_{N,M}^{(z)}\left(\mathcal{A}_{N,M}(n)\right)\, \leq\, \left(p_{N,M}(n,k)\right)^{1/2^k}\, ,\quad 
    \text{ where }\
    p_{N,M}(n,k)\, =\, \mathbb{P}_{N,M}^{(z)}\left(\cap_{t \in \{L,R\}^k} R_{k}^{(t)}(\mathcal{A}_{N,M}(n))\right)\, ,
\end{equation}
where $R_k^{(t)}(\mathcal{A}_{N,M}(n))$ is the event obtained from $\mathcal{A}_{N,M}(n)$ by reflecting $k$ times and choosing the left or right side of each step according to the tiling address $t \in \{L,R\}^k$.
A picture for 2 reflections is shown in Figure \ref{fig:SecondDimer}.

\begin{figure}
$$
\begin{tikzpicture}[xscale=0.3,yscale=0.3,white]
\foreach \y in {0,1,...,5}
{
\foreach \x in {0,1,...,46}
{
\draw (\x,\y) rectangle +(1,1);
}
}
\foreach \y in {0,1,...,5}
{
\foreach \x in {-1,...,46}
{
\draw[gray] (\x,\y) rectangle +(1,1);
}
}
\foreach \x in {11,13,...,19}
{
\draw[black,very thick] (\x,0) -- +(1,0) -- +(1,-1) -- +(2,-1) -- +(2,0);
}
\draw[black,very thick] (21,0) -- +(1,0) -- +(1,-1) -- +(2,-1);
\draw[gray] (23,-1) -- +(0,1);
\foreach \x in {11,13,...,19}
{
\draw[black,very thick] (22-\x,0) -- +(-1,0) -- +(-1,-1) -- +(-2,-1) -- +(-2,0);
}
\draw[black,very thick] (1,0) -- +(-1,0) -- +(-1,-1) -- +(-2,-1);
\draw[gray] (-1,-1) -- +(0,1);
\begin{scope}[xshift=24cm]
\foreach \x in {11,13,...,19}
{
\draw[black,very thick] (\x,0) -- +(1,0) -- +(1,-1) -- +(2,-1) -- +(2,0);
}
\draw[black,very thick] (21,0) -- +(1,0) -- +(1,-1) -- +(2,-1);
\draw[gray] (23,-1) -- +(0,1);
\foreach \x in {11,13,...,19}
{
\draw[black,very thick] (22-\x,0) -- +(-1,0) -- +(-1,-1) -- +(-2,-1) -- +(-2,0);
}
\draw[black,very thick] (1,0) -- +(-1,0) -- +(-1,-1) -- +(-2,-1);
\draw[gray] (-1,-1) -- +(0,1);
\end{scope}
\foreach \x in {1,3,...,9}
{
\filldraw[nearly opaque,blue!50!black,fill=blue!50!white,very thick] (\x,-1) rectangle +(1,2);
}
\foreach \x in {12,14,...,20}
{
\filldraw[nearly opaque,blue!50!black,fill=blue!50!white,very thick] (\x,-1) rectangle +(1,2);
}
\foreach \x in {25,27,...,33}
{
\filldraw[nearly opaque,blue!50!black,fill=blue!50!white,very thick] (\x,-1) rectangle +(1,2);
}
\foreach \x in {36,38,...,44}
{
\filldraw[nearly opaque,blue!50!black,fill=blue!50!white,very thick] (\x,-1) rectangle +(1,2);
}
\foreach \x in {2,4,...,8}
{
\filldraw[nearly opaque,blue!50!black,fill=blue!50!white,very thick] (\x,0) rectangle +(1,2);
}
\foreach \x in {13,15,...,19}
{
\filldraw[nearly opaque,blue!50!black,fill=blue!50!white,very thick] (\x,0) rectangle +(1,2);
}
\foreach \x in {26,28,...,32}
{
\filldraw[nearly opaque,blue!50!black,fill=blue!50!white,very thick] (\x,0) rectangle +(1,2);
}
\foreach \x in {37,39,...,43}
{
\filldraw[nearly opaque,blue!50!black,fill=blue!50!white,very thick] (\x,0) rectangle +(1,2);
}
\foreach \x in {3,5,7}
{
\filldraw[nearly opaque,blue!50!black,fill=blue!50!white,very thick] (\x,1) rectangle +(1,2);
}
\foreach \x in {14,16,18}
{
\filldraw[nearly opaque,blue!50!black,fill=blue!50!white,very thick] (\x,1) rectangle +(1,2);
}
\foreach \x in {27,29,31}
{
\filldraw[nearly opaque,blue!50!black,fill=blue!50!white,very thick] (\x,1) rectangle +(1,2);
}
\foreach \x in {38,40,42}
{
\filldraw[nearly opaque,blue!50!black,fill=blue!50!white,very thick] (\x,1) rectangle +(1,2);
}
\foreach \x in {4,6}
{
\filldraw[nearly opaque,blue!50!black,fill=blue!50!white,very thick] (\x,2) rectangle +(1,2);
}
\foreach \x in {15,17}
{
\filldraw[nearly opaque,blue!50!black,fill=blue!50!white,very thick] (\x,2) rectangle +(1,2);
}
\foreach \x in {28,30}
{
\filldraw[nearly opaque,blue!50!black,fill=blue!50!white,very thick] (\x,2) rectangle +(1,2);
}
\foreach \x in {39,41}
{
\filldraw[nearly opaque,blue!50!black,fill=blue!50!white,very thick] (\x,2) rectangle +(1,2);
}
\foreach \x in {5,16,29,40}
{
\filldraw[nearly opaque,blue!50!black,fill=blue!50!white,very thick] (\x,3) rectangle +(1,2);
}
\begin{scope}[yscale=-1,yshift=-6cm]
% \foreach \y in {0,1,...,5}
% {
% \foreach \x in {0,1,...,46}
% {
% \draw (\x,\y) rectangle +(1,1);
% }
% }
% \foreach \y in {0,1,...,5}
% {
% \foreach \x in {-1,...,46}
% {
% \draw[gray] (\x,\y) rectangle +(1,1);
% }
% }
% \foreach \x in {11,13,...,19}
% {
% \draw[black,very thick] (\x,0) -- +(1,0) -- +(1,-1) -- +(2,-1) -- +(2,0);
% }
% \draw[black,very thick] (21,0) -- +(1,0) -- +(1,-1) -- +(2,-1);
% \draw[gray] (23,-1) -- +(0,1);
% \foreach \x in {11,13,...,19}
% {
% \draw[black,very thick] (22-\x,0) -- +(-1,0) -- +(-1,-1) -- +(-2,-1) -- +(-2,0);
% }
% \draw[black,very thick] (1,0) -- +(-1,0) -- +(-1,-1) -- +(-2,-1);
% \draw[gray] (-1,-1) -- +(0,1);
% \begin{scope}[xshift=24cm]
% \foreach \x in {11,13,...,19}
% {
% \draw[black,very thick] (\x,0) -- +(1,0) -- +(1,-1) -- +(2,-1) -- +(2,0);
% }
% \draw[black,very thick] (21,0) -- +(1,0) -- +(1,-1) -- +(2,-1);
% \draw[gray] (23,-1) -- +(0,1);
% \foreach \x in {11,13,...,19}
% {
% \draw[black,very thick] (22-\x,0) -- +(-1,0) -- +(-1,-1) -- +(-2,-1) -- +(-2,0);
% }
% \draw[black,very thick] (1,0) -- +(-1,0) -- +(-1,-1) -- +(-2,-1);
% \draw[gray] (-1,-1) -- +(0,1);
% \end{scope}
\foreach \x in {1,3,...,9}
{
\filldraw[nearly opaque,blue!50!black,fill=blue!50!white,very thick] (\x,-1) rectangle +(1,2);
}
\foreach \x in {12,14,...,20}
{
\filldraw[nearly opaque,blue!50!black,fill=blue!50!white,very thick] (\x,-1) rectangle +(1,2);
}
\foreach \x in {25,27,...,33}
{
\filldraw[nearly opaque,blue!50!black,fill=blue!50!white,very thick] (\x,-1) rectangle +(1,2);
}
\foreach \x in {36,38,...,44}
{
\filldraw[nearly opaque,blue!50!black,fill=blue!50!white,very thick] (\x,-1) rectangle +(1,2);
}
\foreach \x in {2,4,...,8}
{
\filldraw[nearly opaque,blue!50!black,fill=blue!50!white,very thick] (\x,0) rectangle +(1,2);
}
\foreach \x in {13,15,...,19}
{
\filldraw[nearly opaque,blue!50!black,fill=blue!50!white,very thick] (\x,0) rectangle +(1,2);
}
\foreach \x in {26,28,...,32}
{
\filldraw[nearly opaque,blue!50!black,fill=blue!50!white,very thick] (\x,0) rectangle +(1,2);
}
\foreach \x in {37,39,...,43}
{
\filldraw[nearly opaque,blue!50!black,fill=blue!50!white,very thick] (\x,0) rectangle +(1,2);
}
\foreach \x in {3,5,7}
{
\filldraw[nearly opaque,blue!50!black,fill=blue!50!white,very thick] (\x,1) rectangle +(1,2);
}
\foreach \x in {14,16,18}
{
\filldraw[nearly opaque,blue!50!black,fill=blue!50!white,very thick] (\x,1) rectangle +(1,2);
}
\foreach \x in {27,29,31}
{
\filldraw[nearly opaque,blue!50!black,fill=blue!50!white,very thick] (\x,1) rectangle +(1,2);
}
\foreach \x in {38,40,42}
{
\filldraw[nearly opaque,blue!50!black,fill=blue!50!white,very thick] (\x,1) rectangle +(1,2);
}
\foreach \x in {4,6}
{
\filldraw[nearly opaque,blue!50!black,fill=blue!50!white,very thick] (\x,2) rectangle +(1,2);
}
\foreach \x in {15,17}
{
\filldraw[nearly opaque,blue!50!black,fill=blue!50!white,very thick] (\x,2) rectangle +(1,2);
}
\foreach \x in {28,30}
{
\filldraw[nearly opaque,blue!50!black,fill=blue!50!white,very thick] (\x,2) rectangle +(1,2);
}
\foreach \x in {39,41}
{
\filldraw[nearly opaque,blue!50!black,fill=blue!50!white,very thick] (\x,2) rectangle +(1,2);
}
\foreach \x in {5,16,29,40}
{
\filldraw[nearly opaque,blue!50!black,fill=blue!50!white,very thick] (\x,3) rectangle +(1,2);
}
\end{scope}
\end{tikzpicture}
$$
\caption{An example of ``bow-tie'' shaped regions obtained from reflecting the triangles about height $m\leq n$.}
    \label{fig:ThirdDimer}
\end{figure}
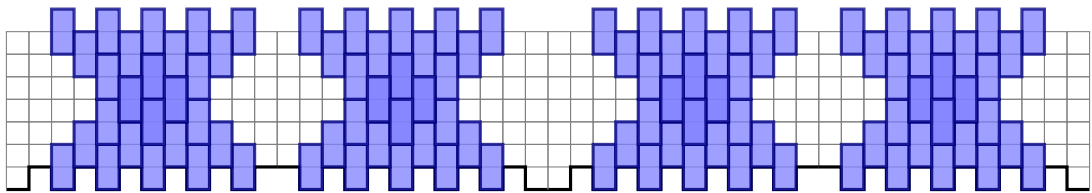

Secondly, one may also reflect in the vertical direction.
If one reflects in a horizontal line below the top of the frozen triangles (in Figure \ref{fig:SecondDimer})
then this means that one obtains frozen bow-tie shaped regions.
We have tried to display an example in Figure \ref{fig:ThirdDimer}.

In particular, if one reflects at a height of $m<n$ then the density of unfrozen squares in the rectangle $\{0,\dots,n-1\} \times \{0,\dots,2m-1\}$
is $2m(m-1)$-divided-by-$2mn$. So the density is less than $m/n$. If we choose a fixed, but small $\epsilon$ so that $m\leq n\epsilon$,
then we obtain a density less than $\epsilon$ for free dimers.
But this means that the probability is small by an argument involving instability at low entropy.

Namely, one may obtain a reference state with a fixed positive entropy in various ways. Let us mention one way.
Suppose that $\ell$ is a positive even number.
Consider decomposing $\{1,\dots,N\}$ into $\lfloor N/\ell \rfloor$ blocks of the form $\{(k-1)\ell + 1,\dots,k \ell\}$ for $k \in \{1,\dots,\lfloor N/\ell \rfloor\}$
(and with a small remainder of sites at the end if $N$ is not an integer multiple of $\ell$).
For each block, consider the $\ell/2$ even sites.
For each horizontal row of squares at even heights (so only half the rows), on each block, choose one of the $\ell/2$ sites uniformly at random,
IID.
Choose that square to have D state and hence the square above it on a row of odd height to be U.
All other squares are filled by horizontal dimers, with state L if the site has x-coordinate even and R if the square has $x$-coordinate odd.

This way, we have $\ln(\text{Number})/(MN\ln(2))$ is approximately $\ln(\ell/2)/(2\ell \ln(2))$.
This is quite small for large $\ell$.
But it is fixed, and not zero.
On the other hand, by fixing $\ell$, and then choosing $\epsilon\approx m/n$ sufficiently small in comparison, we can see that the 
probability of the intersection of all the tile events after maximal reflections is exponentially small of the form $e^{-\eta MN}$ where $\eta$
does depend on our choices for $\ell$ and $\epsilon$.
But the number of roots we have to take due to the reflection positive bound is less than this by a multiple $mn$. So we do obtain an upper bound on 
the EFP of the form $e{-Cmn}$ which is $e^{-\epsilon n^2}$.

This is all assuming that $M$ is sufficiently large that we may take $m=\lfloor \epsilon n \rfloor$ without exceeding $M$.
(There are technicalities in ``reflection positivity'' about remainders when $N$ and $M$ are not perfect powers of 2, but 
those details do work out. We refer to \cite{CrawfordNgStarr} for an example where such details are done.)
If $M$ is not sufficiently large then the same type of argument can be run, but then $mn$ is replaced by $nM$. That explains why the bound
has the form $e^{-c n \min(n,M)}$ in  Theorem \ref{thm:main}.

\begin{figure}
$$
\begin{tikzpicture}[xscale=0.3,yscale=0.3,white]
\foreach \y in {0,1,...,5}
{
\foreach \x in {0,1,...,46}
{
\draw (\x,\y) rectangle +(1,1);
}
}
\foreach \y in {0,1,...,5}
{
\foreach \x in {-1,...,46}
{
\draw[gray] (\x,\y) rectangle +(1,1);
}
}
\foreach \x in {11,13,...,19}
{
\draw[black,very thick] (\x,0) -- +(1,0) -- +(1,-1) -- +(2,-1) -- +(2,0);
}
\draw[black,very thick] (21,0) -- +(1,0) -- +(1,-1) -- +(2,-1);
\draw[gray] (23,-1) -- +(0,1);
\foreach \x in {11,13,...,19}
{
\draw[black,very thick] (22-\x,0) -- +(-1,0) -- +(-1,-1) -- +(-2,-1) -- +(-2,0);
}
\draw[black,very thick] (1,0) -- +(-1,0) -- +(-1,-1) -- +(-2,-1);
\draw[gray] (-1,-1) -- +(0,1);
\begin{scope}[xshift=24cm]
\foreach \x in {11,13,...,19}
{
\draw[black,very thick] (\x,0) -- +(1,0) -- +(1,-1) -- +(2,-1) -- +(2,0);
}
\draw[black,very thick] (21,0) -- +(1,0) -- +(1,-1) -- +(2,-1);
\draw[gray] (23,-1) -- +(0,1);
\foreach \x in {11,13,...,19}
{
\draw[black,very thick] (22-\x,0) -- +(-1,0) -- +(-1,-1) -- +(-2,-1) -- +(-2,0);
}
\draw[black,very thick] (1,0) -- +(-1,0) -- +(-1,-1) -- +(-2,-1);
\draw[gray] (-1,-1) -- +(0,1);
\end{scope}
\foreach \x in {1,3,...,9}
{
\filldraw[nearly opaque,blue!50!black,fill=blue!50!white,very thick] (\x,-1) rectangle +(1,2);
}
\foreach \x in {12,14,...,20}
{
\filldraw[nearly opaque,blue!50!black,fill=blue!50!white,very thick] (\x,-1) rectangle +(1,2);
}
\foreach \x in {25,27,...,33}
{
\filldraw[nearly opaque,blue!50!black,fill=blue!50!white,very thick] (\x,-1) rectangle +(1,2);
}
\foreach \x in {36,38,...,44}
{
\filldraw[nearly opaque,blue!50!black,fill=blue!50!white,very thick] (\x,-1) rectangle +(1,2);
}
\foreach \x in {2,4,...,8}
{
\filldraw[nearly opaque,blue!50!black,fill=blue!50!white,very thick] (\x,0) rectangle +(1,2);
}
\foreach \x in {13,15,...,19}
{
\filldraw[nearly opaque,blue!50!black,fill=blue!50!white,very thick] (\x,0) rectangle +(1,2);
}
\foreach \x in {26,28,...,32}
{
\filldraw[nearly opaque,blue!50!black,fill=blue!50!white,very thick] (\x,0) rectangle +(1,2);
}
\foreach \x in {37,39,...,43}
{
\filldraw[nearly opaque,blue!50!black,fill=blue!50!white,very thick] (\x,0) rectangle +(1,2);
}
\foreach \x in {3,5,7}
{
\filldraw[nearly opaque,blue!50!black,fill=blue!50!white,very thick] (\x,1) rectangle +(1,2);
}
\foreach \x in {14,16,18}
{
\filldraw[nearly opaque,blue!50!black,fill=blue!50!white,very thick] (\x,1) rectangle +(1,2);
}
\foreach \x in {27,29,31}
{
\filldraw[nearly opaque,blue!50!black,fill=blue!50!white,very thick] (\x,1) rectangle +(1,2);
}
\foreach \x in {38,40,42}
{
\filldraw[nearly opaque,blue!50!black,fill=blue!50!white,very thick] (\x,1) rectangle +(1,2);
}
\foreach \x in {4,6}
{
\filldraw[nearly opaque,blue!50!black,fill=blue!50!white,very thick] (\x,2) rectangle +(1,2);
}
\foreach \x in {15,17}
{
\filldraw[nearly opaque,blue!50!black,fill=blue!50!white,very thick] (\x,2) rectangle +(1,2);
}
\foreach \x in {28,30}
{
\filldraw[nearly opaque,blue!50!black,fill=blue!50!white,very thick] (\x,2) rectangle +(1,2);
}
\foreach \x in {39,41}
{
\filldraw[nearly opaque,blue!50!black,fill=blue!50!white,very thick] (\x,2) rectangle +(1,2);
}
\foreach \x in {5,16,29,40}
{
\filldraw[nearly opaque,blue!50!black,fill=blue!50!white,very thick] (\x,3) rectangle +(1,2);
}
\end{tikzpicture}
$$
\caption{In this figure we have performed 2 reflections, as in the method of reflection positivity. We have
also indicated the other {\em forced} dimers, which are forced by the boundary condition on the first row, which
shows that the boundary condition really does affect at least an order of $n^2$ other vertices.}
    \label{fig:SecondDimer}
\end{figure}
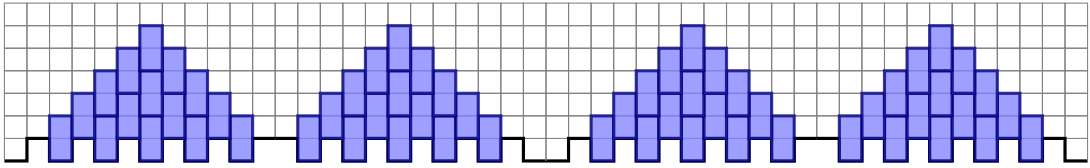

% \section*{Acknowledgments}
% S.S.~is grateful to Paul Jung for some useful conversations.
% We are grateful to Nayantara Bhatnagar for discussing her result in \cite{BhatnagarPeled} with us, and especially for pointing out the 
% importance of obtaining quantitative versions of a weak law. 
% We are also appreciative of the 
% helpful corrections and improvements of an anonymous referee. 
% We have indicated some of these in remarks in the paper.

\baselineskip=12pt
\bibliographystyle{plain}

\noindent
\underline{Contact:}
\texttt{slstarr@uab.edu}

\end{document}